\documentclass[sigconf]{acmart}

\usepackage{booktabs} % For formal tables

\usepackage{listings}
\graphicspath{{figures/}}
\DeclareGraphicsExtensions{.pdf,.jpg,.png,.eps}
\newcommand{\papas}{\texorpdfstring{P\lowercase{a}P\lowercase{a}S}\ }
\usepackage{balance}

\copyrightyear{2018}
\acmYear{2018}
\setcopyright{acmcopyright}
\acmConference[PEARC '18]{Practice and Experience in Advanced Research Computing}{July 22--26, 2018}{Pittsburgh, PA, USA}
\acmBooktitle{PEARC '18: Practice and Experience in Advanced Research Computing, July 22--26, 2018, Pittsburgh, PA, USA}
\acmPrice{15.00}
\acmDOI{10.1145/3219104.3229289}
\acmISBN{978-1-4503-6446-1/18/07}

\begin{document}

\title[\papas: A Generic Framework for Parallel Parameter Studies]{\papas: A Portable, Lightweight, and Generic Framework for Parallel Parameter Studies}

\author{Eduardo Ponce}
\orcid{0002-8854-9043}
\affiliation{%
  \department{Electrical Engineering \& Computer Science}
  \institution{The University of Tennessee}
  \streetaddress{Circle Park Dr.}
  \city{Knoxville}
  \state{TN}
  \postcode{37996}
}
\email{eponcemo@utk.edu}

\author{Brittany Stephenson}
\affiliation{%
  \department{Mathematics}
  \institution{The University of Tennessee}
  \streetaddress{Circle Park Dr.}
  \city{Knoxville}
  \state{TN}
  \postcode{37996}
}
\email{bsteph13@vols.utk.edu}

\author{Suzanne Lenhart}
\orcid{0002-6898-5796}
\affiliation{%
  \department{Mathematics}
  \institution{The University of Tennessee}
  \streetaddress{Circle Park Dr.}
  \city{Knoxville}
  \state{TN}
  \postcode{37996}
}
\email{lenhart@math.utk.edu}

\author{Judy Day}
\orcid{0002-6544-1355}
\affiliation{
  \department[1]{Mathematics}
  \department[0]{Electrical Engineering \& Computer Science}
  \institution{The University of Tennessee}
  \streetaddress{Circle Park Dr.}
  \city{Knoxville}
  \state{TN}
  \postcode{37996}
}
\email{judyday@utk.edu}

\author{Gregory D. Peterson}
\orcid{0002-0875-5278}
\affiliation{%
  \department{Electrical Engineering \& Computer Science}
  \institution{The University of Tennessee}
  \streetaddress{Circle Park Dr.}
  \city{Knoxville}
  \state{TN}
  \postcode{37996}
}
\email{gdp@utk.edu}

% The default list of authors is too long for headers.
\renewcommand{\shortauthors}{E. Ponce et al.}
\renewcommand{\authors}{E. Ponce, B. Stephenson, S. Lenhart, J. Day, and G.D. Peterson}

\begin{abstract}
The current landscape of scientific research is widely based on
modeling and simulation, typically with complexity in the simulation's
flow of execution and parameterization properties.
Execution flows are not necessarily
straightforward since they may need multiple processing tasks and iterations.
Furthermore, parameter and performance studies are common approaches used
to characterize a simulation, often requiring traversal of a large parameter space.
High-performance computers offer practical resources
at the expense of users handling the setup, submission, and management of jobs.
This work presents the design of \papas, a portable, lightweight, and generic
workflow framework for conducting parallel parameter and performance studies.
Workflows are defined using parameter files based on keyword-value pairs syntax,
thus removing from the user the overhead of creating complex scripts to manage the workflow.
A parameter set consists of any combination of environment variables,
files, partial file contents, and command line arguments.
\papas\ is being developed in Python 3 with support for distributed parallelization
using SSH, batch systems, and C++ MPI.
The \papas\ framework will run as user processes,
and can be used in single/multi-node and multi-tenant computing systems.
An example simulation using the BehaviorSpace tool from NetLogo and a matrix
multiply using OpenMP are presented as parameter and performance studies, respectively.
The results demonstrate that the \papas\ framework offers a simple method for defining
and managing parameter studies, while increasing resource utilization.

\end{abstract}

%
% The code below should be generated by the tool at
% http://dl.acm.org/ccs.cfm
% Please copy and paste the code instead of the example below.
%



\maketitle

\section{Introduction} \label{intro}

Computational approaches, such as modeling and simulation, are widely used
to find information and patterns otherwise not readily available.
These applications are often complex due to the sheer size of the parameter
space and long run times~\cite{walker2007challengesparametersweep}.
Moreover, applications may consist of compound basic workflow structures (\textit{e.g.} process, pipeline, data distribution, data aggregation, and data
redistribution~\cite{bharathi2008characterizationworkflows}).
As a consequence, testing, monitoring, and validating workflows is not trivial because parameters may come from
disparate sources (\textit{e.g.}, command line arguments, environment variables, files, or a combination of these).

High-performance clusters and grid systems are practical for
performing parameter studies due to their large collection of processors
and storage resources~\cite{krol2016scalarm,deelman2018scientificworkflows,tilson2008workflowsperformanceevaluation}, although
local computers can also be used due to the
advancement of graphic processors and other accelerators~\cite{ino2014parametersweepgpu}.
The setup, submission, and orchestration of such
jobs in computing clusters may be a challenge, particularly to
non-programmers or novice users
for conducting parameter studies in a parallel or distributed fashion~\cite{devivo2001comparisonparametertools,prodan2004zenturio}.
Previous work has evaluated static and dynamic scheduling algorithms for managing workflow structures efficiently in cluster and grid systems~\cite{yu2008gridschedulingalgorithms,smanchat2009schedulingparameterworkflow,buyya2002economicmodelsgrid}.
Clearly, scientific research has benefited from systematic
parameter studies as a means to find an optimal or reasonable set of parameters~\cite{sedlmair2014visualparameterspace,fey2004frameworkparameterstudies,tan2005paramstudyfiber}.

This work presents the ongoing effort of designing \papas, an easy-to-use Python 3 framework for describing and executing parameter and performance studies in local and cluster computers.
\papas\ serves as a lightweight workflow manager configured via a simple keyword/value workflow language.
In Section~\ref{related} we present examples of existing tools for parameter studies.
A representative scenario of job's execution in cluster systems is used as motivation in Section~\ref{motivation},
contributions of \papas\ are included here as well.
Section~\ref{design} gives a general overview of \papas\ architectural design,
followed by a description of its workflow language, Section~\ref{papas_lang}.
As case studies, a parameter study of a multi-agent NetLogo model~\cite{wilensky2015agentmodeling} is
found in Section~\ref{study_netlogo}, and
a parameter study of a matrix multiply using OpenMP threading library is shown in Section~\ref{study_matmul}.
The remainder of the manuscript summarizes the work and describes several enhancements that
will make \papas\ interact well with existing workflow management tools.

\section{Related Work} \label{related}

There are numerous available tools and frameworks for running parameter studies
in cluster and grid-enabled systems.
Commonly, these workflow management systems need to be installed by
administrators as system-wide software, and include web interfaces for user interaction~\cite{ries2010comsolgrid,wolstencroft2013tavernasuite,volkov2015genericparametersweep,smirnov2016distributedresourceseverest}.
Also, some of these tools have a high number of
modules interacting with one another which makes the system not ideal
for either novice computer users or simple jobs.
It is worth noting that workflow management systems have been studied since the advent
of the Internet and are widely used.
Several mature projects are available, for example, Taverna~\cite{missier2010taverna}, Pegasus~\cite{deelman2015pegasus}, and Nimrod/K~\cite{abramson2008nimrodk}.

In the area of parameter studies, OACIS~\cite{murase2017oacis} is a management framework for exploring parameter
spaces. It provides a web interface for submitting and monitoring jobs sent to remote system.
A limitation is that the simulation can use inputs provided from the command line or a file.
OpenMOLE~\cite{reuillon2013openmole} is a framework that supports
distributed NetLogo~\cite{wilensky2008netlogo} runs, at the expense that a configuration file based on a domain
specific language is generated by the user.
This configuration file includes parameters and tasks, and controls the job's distribution.
A simpler workflow application, Snakemake, runs as a single user's program and is written in Python~\cite{koster2012snakemake}.
Snakemake is based on GNU Make syntax and infers task dependencies from files dependencies.

\section{\papas\ Motivation} \label{motivation}

\begin{figure}
    \includegraphics[scale=1.10]{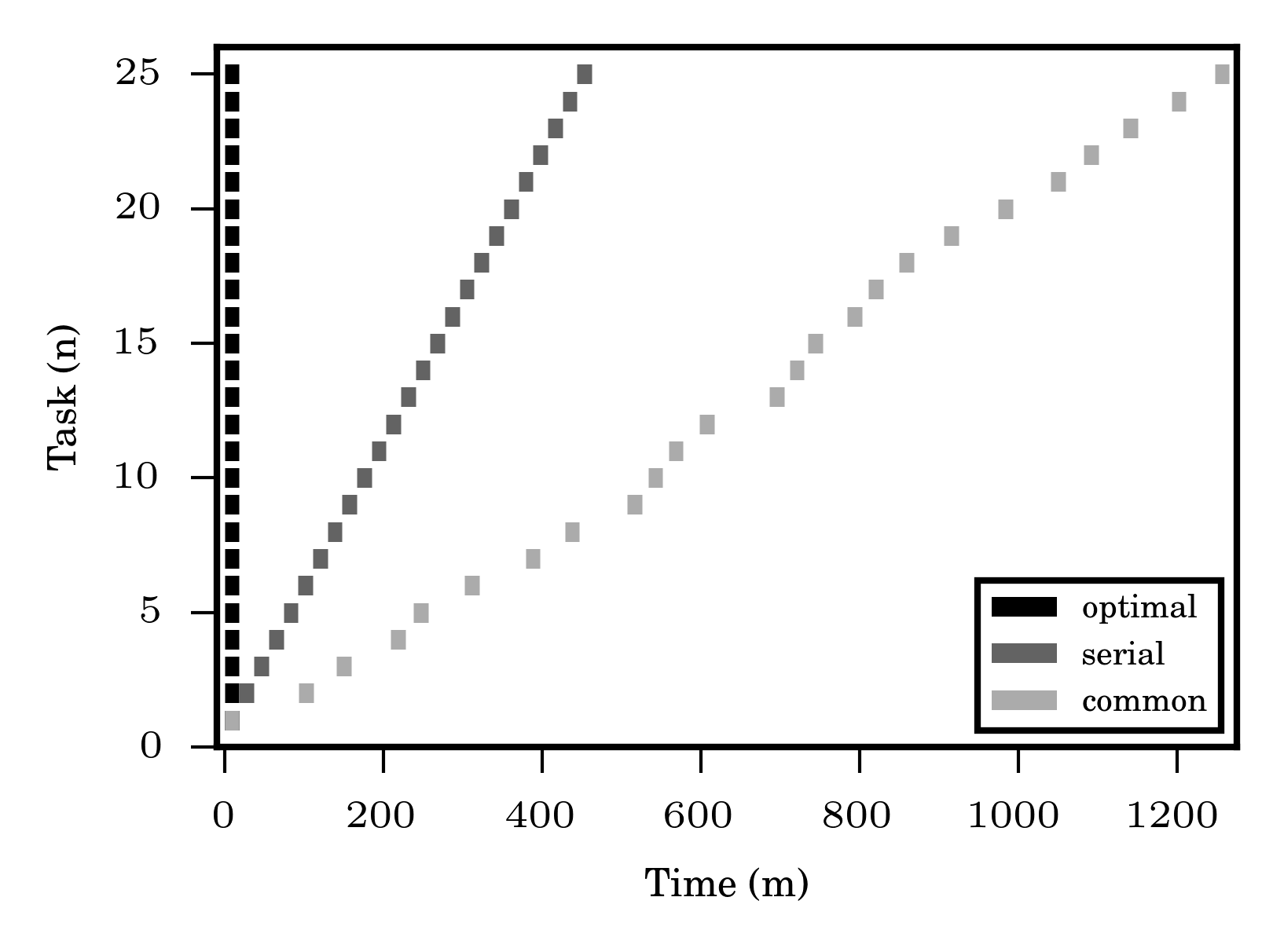}
    \caption{Representation of execution behavior of 25 jobs running in a managed multi-user cluster under different forms of submission, scheduling, and cluster activity. For each submission form, all jobs are submitted simultaneously. The {\itshape optimal} scenario corresponds to submitting 25 jobs to a cluster with at least 25 available compute nodes. Every job starts and ends at the same time. The {\itshape serial} case occurs when the scheduler decides to
    run one job at a time, without delays between the end and start of consecutive tasks. If the cluster activity is high or the scheduler is not fair enough, consecutive tasks will have different delays in between and a {\itshape common} scenario takes place.}
    \label{fig:cluster_stats}
\end{figure}

High-performance clusters are desired to have a high utilization activity to
quantify a positive return-of-investment~\cite{fulton2017xdmod,simakov2015applicationkernels}, since these are costly systems to build and maintain.
Software monitoring tools, \textit{e.g.,} XDMoD~\cite{furlani2013xdmodxsede,palmer2015openxdmod}, are valuable for gathering vast amounts of performance metrics used to improved large-scale multi-user systems.
The execution behavior of jobs is affected by the submission order, scheduling heuristics~\cite{zaharia2010delay}, and utilization rate. Figure~\ref{fig:cluster_stats} presents cases representable of the start and stop times of 25 jobs.
For every task the scheduler has to handle the start and stop actions, this overhead can be reduced
if multiple user jobs are batched together into a single cluster job.
Parameter studies require the execution of many application's instances,
this is a combinatorial optimization problem~\cite{blum2003metaheuristics}.

The motivation of this work is to provide a simple
methodology for performing parameter studies for general application classes, while improving
a system's utilization and reducing overall completion time.
\papas\ is a versatile framework for describing parameter and performance studies using flexible configuration files.
Section~\ref{combinatorial} exposes the combinatorial approach used in \papas\ to enumerate all possible
unique workflow instances.
The primary contributions of the \papas\ framework are:
\begin{itemize}
    \item Deploying a user-space tool for expressing workflows targeted at parameter and performance studies,
          with no administrator or system-wide installations
    \item Expressiveness of parameter study workflows using common text formats
          (\textit{i.e.,} YAML, JSON, INI), preventing users to write complex scripts
    \item Combinations of parameters can be a mix of command line arguments,
          environment variables, files, and simple regular expressions for file contents, and
    \item Support for batching user jobs as a single cluster job using MPI library
\end{itemize}

Another motivation for the \papas\ framework is its applicability for
evaluating machine learning and natural language processing algorithms ~\cite{mayer2018textanalytics}.
Due to the numerous set of tunable hyperparameters ~\cite{witt2017hyperparameters}
and the breadth of machine learning toolkits available, both parameter and performance studies are labor-intensive.
\papas\ framework can provide immediate benefits to such scenarios.

\section{\papas\ Framework Design} \label{design}

\begin{figure*}
    \includegraphics[scale=0.25]{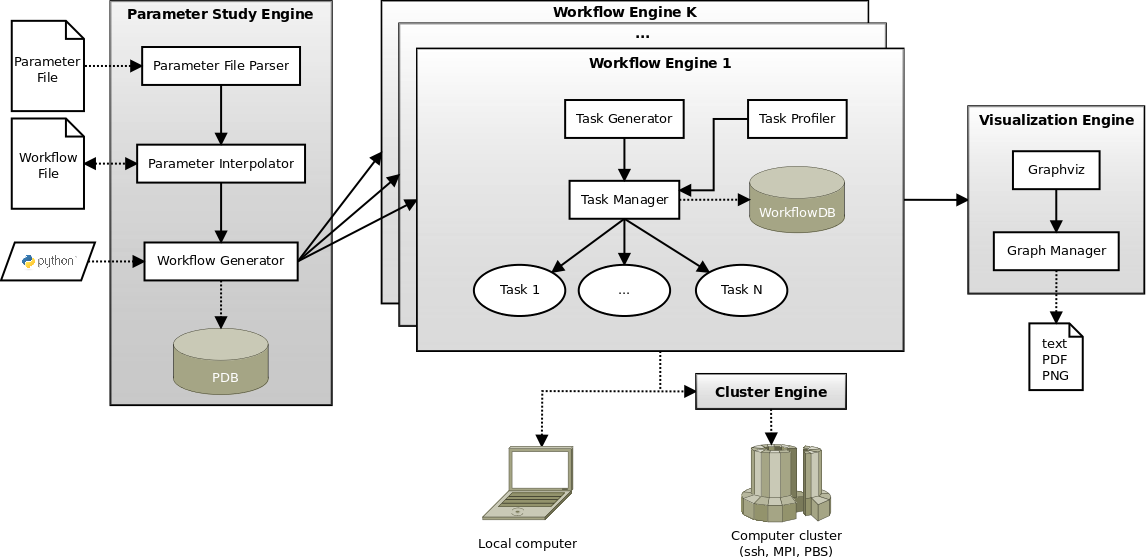}
    \caption{\papas\ architecture consists of four principal and modular engines: (1) parameter study, (2) workflow, (3) cluster, and (4) visualization. User interacts with the parameter study engine using parameter files or the Python 3 API. A workflow engine manages the execution of workflows as well as gathering profiling and provenance information. The visualization engine serves a visual aid for validating a parameter study and for visual monitoring.}
    \label{fig:papas_arch}
\end{figure*}

Previous works have shown that parameter studies require
several operations for effectively managing the workflows:
value propagation via dependency graphs, orchestration, I/O management,
monitoring, provenance, visualization, on-line feedback, and
others~\cite{shi2006scheduling,devivo2001comparisonparametertools,walker2007challengesparametersweep}.
The \papas\ framework is a collection of modular systems, each with unique
functionality and independent interfaces.
Figure~\ref{fig:papas_arch} presents the overall architectural design of \papas.
The primary system components are the
parameter study, workflow, cluster, and visualization
engines.

\subsection{Parameter Study Engine} \label{parameter_engine}
A parameter study represents a set of workflows to be executed,
where a workflow corresponds to an instance having a unique parameter combination.
Users write a parameter file using a keyword-based workflow description language
which is described in Section~\ref{papas_lang}.
A workflow's description can be divided across multiple parameter files; this
allows composition and re-usability of task configurations.
Parameter files follow either YAML, JSON, or INI-like data serialization formats
with minor constraints.
The processing of these files consists of a parsing and syntax validation step,
followed by string interpolation for parameters that were specified with
multiple values.
The operation of interpolation identifies all the possible unique parameter
combinations and forwards this information to a workflow generator which in
turn spawns a workflow engine instance per combination.
Parameter study configurations are stored in a file database as part
of the monitoring activity.
\papas\ provides checkpoint-restart functionality in case of fault or a deliberate
pause/stop operation.
A parameter study's state can be saved in a workflow file and reloaded at a later time.
Another method of defining a parameter study is through the workflow
generator Python 3 interface.
This mechanism adds the hooks to embed \papas\ as a task of a larger user-defined
workflow.

\subsection{Workflow Engine} \label{workflow_engine}
Workflow engines are a core component as they orchestrate the
execution of workflow instances.
The task generator takes a workflow description and constructs
a directed acyclic graph (DAG) where nodes correspond to indivisible tasks.
A task manager controls the scheduling and monitoring of tasks.
\papas\ runs easily on a local laptop or workstation.
For cluster systems, workflow tasks are delegated to the
cluster engine component.
Several factors affect scheduling heuristics such as task dependencies,
availability and capability of computing resources,
and the application(s) behavior.
A task profiler measures each task's runtime, but
currently this only serves as performance feedback to the user.
Workflow engine actions, task/workflow statistics, and logs are stored in a per-workflow
file storage database; this information is later used to include
provenance details at either workflow completion or a checkpoint.
A visualization engine enables access to a view of the workflow's DAG.
The workflow engine communicates the progression of states to the
visualization engine.

\subsection{Cluster Engine} \label{cluster_engine}
The cluster engine is a component that serves as an interface for both
managed and unmanaged computer clusters.
A managed cluster is assumed to be used concurrently by multiple users
and makes use of a batch system (\textit{e.g.}, PBS, SGE),
while an unmanaged cluster is mostly single-user and has a SSH setup.
For managed clusters, the common approach is to submit a single task
per batch job.
Single task submissions are mainly applicable for applications that achieve
a high-utilization of computing resources or have long execution times, and
adding concurrent task executions hinder performance.
For single-node and single-core applications, submitting a
large number of jobs to a multi-tenant system may not necessarily be the best
approach.
\papas\ workflow and cluster engines enable grouping intra/inter-workflow
tasks as a single batch job.
The main mechanism for grouping tasks as single jobs is using a C++ MPI task dispatcher.
In some cases, task grouping increases the cluster's utilization efficiency, reduces
batch/scheduling operations, and improves turnaround time of jobs.
Section~\ref{study_netlogo} presents a case study portraying these effects.

\subsection{Visualization Engine} \label{viz_engine}
The DAGs generated by the workflow engine are used to construct
visual graphs of the overall workflow as well as the current state of
the processing.
\papas\ utilizes a wrapper over PyGraphviz~\cite{hagbergpygraphviz} to build
and update graphs on-demand.
A workflow visualization can be viewed and exported in text or common image formats.
This capability can also be enabled as a validation method of the parameter study
configuration prior to any execution taking place.

\section{\papas\ Workflow Description Language} \label{papas_lang}

This section describes the workflow description language (WDL) specification used
by the \papas\ framework.
The \papas\ WDL consists of a set of keywords that
can describe individual tasks, task dependencies, parameter sets,
and general configurations.
This is in contrast to the common
description methods for workflows and parameter studies:
parametric modeling languages~\cite{abramson2000nimrodg},
DAG languages~\cite{farkas2011pgradeportal},
XML~\cite{deelman2015pegasus},
task data flow~\cite{wozniak2013swift}, declarative languages~\cite{reuillon2010declarativetaskopenmole},
libraries extending existing programming languages~\cite{bergstra2013hyperopt}, template
systems~\cite{lorca2011gridwayparameter,casanova2000apples}, graphical languages~\cite{yarrow2000uiparamstudy}, UML
diagrams~\cite{dumas2001umlworkflowlanguage}, GNU Make-based~\cite{koster2012snakemake}, test systems~\cite{yu2013acts},
and others~\cite{van2005yawl}.
An advantage of using a keyword-value WDL is that it can impose
stricter constraints to reduce complex and convoluted expressions that are allowed
on other WDLs, as a driving philosophy of \papas\ is to be simple and accessible to support non-programmers.

\papas's WDL is based on a mix of lists and associative structures. As a consequence,
it is serializable and can be converted to common human-readable formats such as YAML, JSON, and INI.
Workflow descriptions are transformed into a common internal format.
The following is the general specification of rules for configuring parameter studies using YAML format.

\begin{itemize}
    \item A parameter study consists of tasks (or sections), identified
          by a {\itshape task} (or {\itshape section}) as the only key, and followed
          by up to two levels of {\itshape keyword-value} entries. That is, the first set of
          values can themselves be a pair of {\itshape keyword-value} entries.
    \item The delimiter for {\itshape keyword-value} entries is the colon character.
	\item Indentation, tab or whitespace, is used to make a {\itshape value} pertain to a
          particular {\itshape keyword}.
    \item A single-line comment is a line that starts with a pound or hash symbol (\#).
    \item A {\itshape keyword} can be specified using any alphanumeric character.
    \item All {\itshape keywords} are parsed as strings and {\itshape values} are inferred
          from written format.
    \item {\itshape Keywords} that are not predefined are considered as a user-defined
          {\itshape keywords} and can be used in value interpolations.
    \item Ranges with a step size are supported for numerical values using the notation {\itshape start:step:end}.
    \item A {\itshape task} is identified by the {\itshape command} keyword.
    \item Value interpolation uses a flat associative array syntax.
	\item Intra-task interpolation using \$\{$\dotsc$\} syntax is allowed using {\itshape values} from both entry
          levels (e.g., \$\{{\itshape keyword}\} and \$\{{\itshape keyword:value}\}).
	\item Inter-task interpolation using \$\{$\dotsc$\} syntax is allowed using {\itshape values} from both entry
          levels (\textit{e.g.}, \$\{{\itshape task:keyword}\} and \$\{{\itshape task:keyword:value}\}).
\end{itemize}

The list below presents a list of common keywords corresponding to \papas\ WDL:
\begin{itemize}
    \item {\bfseries command} -- string representing the command line to run
    \item {\bfseries name} -- string describing the task
    \item {\bfseries environ} -- dictionary of environment variables where
          {\itshape keywords} are the actual names of the environment variables.
    \item {\bfseries after} -- list of tasks dependencies, prerequisites
    \item {\bfseries infiles} -- dictionary of input files, {\itshape keywords} are arbitrary
    \item {\bfseries outfiles} -- dictionary of output files, {\itshape keywords} are arbitrary
    \item {\bfseries substitute} -- used for interpolation of partial file contents. Expects a {\itshape
          keyword/value} pair where {\itshape keyword} is a Python 3 regular expression
          and {\itshape value} is a list of strings to be used instead.
    \item {\bfseries parallel} -- mode to use for parallelism, (\textit{e.g.}, ssh, MPI)
    \item {\bfseries batch} -- batch system of cluster (\textit{e.g.}, PBS)
    \item {\bfseries nnodes} -- number of nodes to use for a cluster job
    \item {\bfseries ppnode} -- number of task processes to run per nodes
    \item {\bfseries hosts} -- hostnames or IP addresses of compute nodes
    \item {\bfseries fixed} -- list of parameters to be
          fixed. All of these parameters need to have the
          same number of values to allow ordered one-to-one mappings.
    \item {\bfseries sampling} -- samples a subset of the parameter space based
          on a given distribution (uniform, random).
\end{itemize}

\subsection{Parameter combinatorial approach} \label{combinatorial}

A key aspect of \papas\ framework is its approach for expressing parameter
combinations easily while being general enough for most parameter and performance
studies.
Every parameter and its values are implicitly used to generate the Cartesian
product of parametric combinations. Each unique combination of parameters
represents a workflow to be executed.

Parameters have a unique name and are allowed to be multi-valued.
Consider a set $\mathcal{P}$ of $m$ parameters,
$\mathcal{P} = \left \{P_1, P_2, \dotsc, P_m \right \}$,
where $P_i$ is a parameter with $N_i$ possible values and $P_{i,j}$ corresponds to
the $j^{th}$ value of $P_i$.
A total of $N_W = \prod_{i=1}^{m} N_i$ workflows are generated automatically by the
\papas\ workflow engine.
Then, this workflow set, $\mathcal{W}$, is defined as
\begin{displaymath}
    \mathcal{W} = \left \{P_1 \times P_2 \times \cdots \times P_m \right \}
\end{displaymath}
\begin{displaymath}
    P_i \times P_j = \left \{ (a, b) \mid a \in P_i,\ b \in P_j\ and\ i,j \in \left \{ 1, 2, \dotsc, m \right \}\ and\ i \neq j \right \}
\end{displaymath}
\begin{displaymath}
    P_i \times P_j = P_j \times P_i
\end{displaymath}
A \papas\ workflow is an instance of $\mathcal{W}$.
Programmatically, this can be implemented as $m$ nested loop structures.
In some cases, the Cartesian product of parameters is either not desired
or too large to run all workflow combinations in a reasonable amount of time.
\papas\ utilizes the keywords {\itshape fixed} and {\itshape sampling} to control the combinatorial
set of workflows, $\mathcal{T}$. Parameters listed in the {\itshape fixed} set
need to have the same number of values to allow one-to-one mappings between
each other.
Workflows will be generated from the Cartesian product of parameters
not listed as {\itshape fixed} combined with a single set of values made from
the ordered values of {\itshape fixed} parameters.
Multiple {\itshape fixed} statements are allowed in a \papas\ parameter file,
this further generalizes combinations and can be used to specify constant
single-valued parameters.
Programmatically, this can be implemented by moving all the {\itshape fixed} parameters
into the outermost loop structures (grouped by same {\itshape fixed} clauses).
For example, consider the total number of workflows generated for $m$ parameters with
$P_2$ and $P_3$ listed in the same {\itshape fixed} clause.
Then,
\begin{displaymath}
    W_1 = \left \{P_1 \times P_4 \times \cdots \times P_m \right \}
\end{displaymath}
\begin{displaymath}
    W_2 = \left \{(P_{2,j}, P_{3,j}) \mid j \in 1, 2, \dotsc, N_2\ and\ N_2 = N_3 \right \}
\end{displaymath}
\begin{displaymath}
    \mathcal{W} = \left \{W_1 \times W_2 \right \}
\end{displaymath}
where $W_i$ represents an incomplete subset of workflows (parameters are
missing).

\section{Parameter sweep: Distributed NetLogo model} \label{study_netlogo}

\begin{figure}
    \includegraphics[scale=1.10]{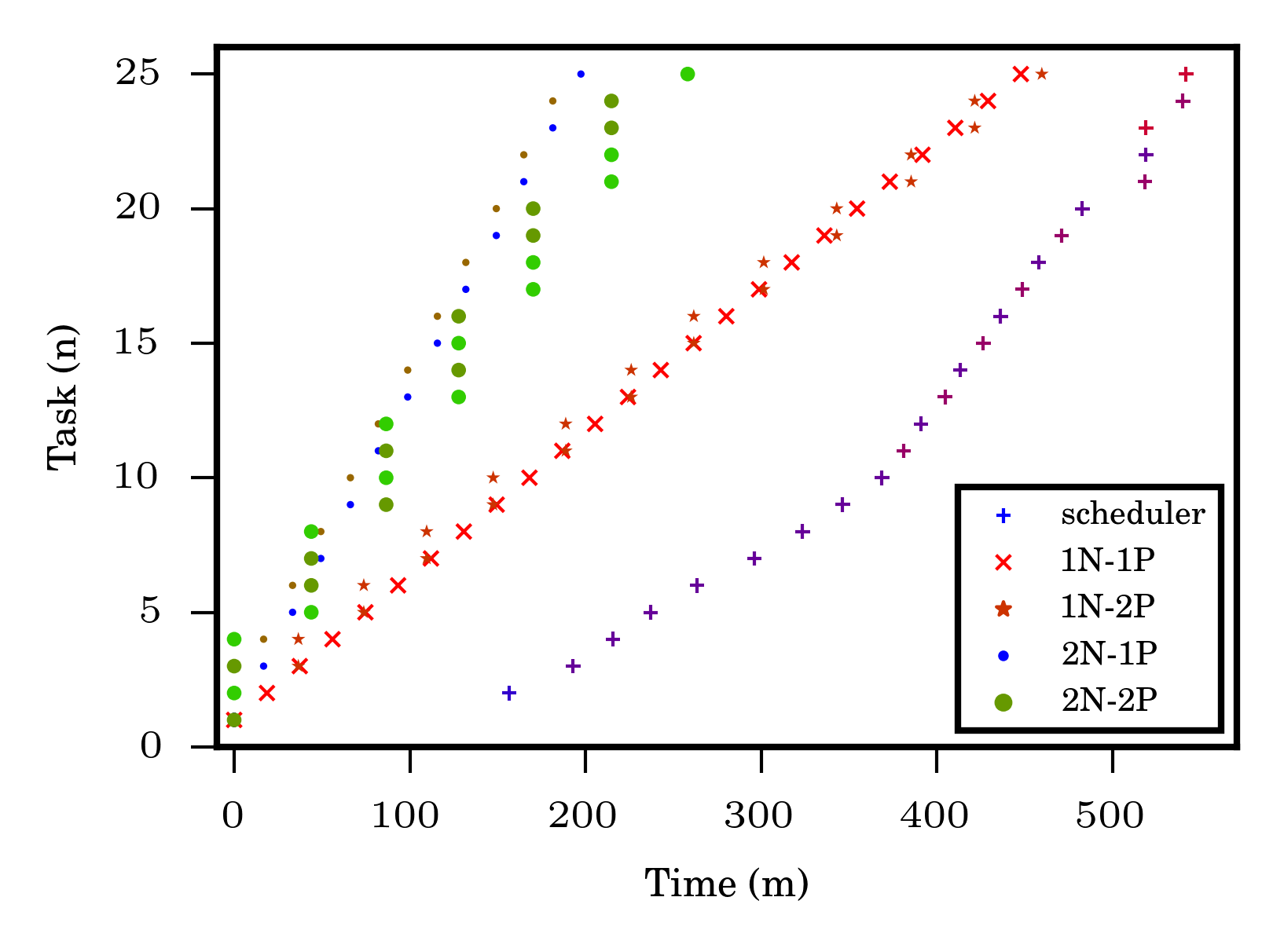}
    \caption{Initial execution behavior of 25 NetLogo simulations using different grouping schemes in terms of compute nodes (N) and number of MPI processes per node (P). Time begins as soon as a job started execution. Note that the scheduler start times have the greater variability.}
    \label{fig:netlogo_begin}
\end{figure}

\begin{figure}
    \includegraphics[scale=1.10]{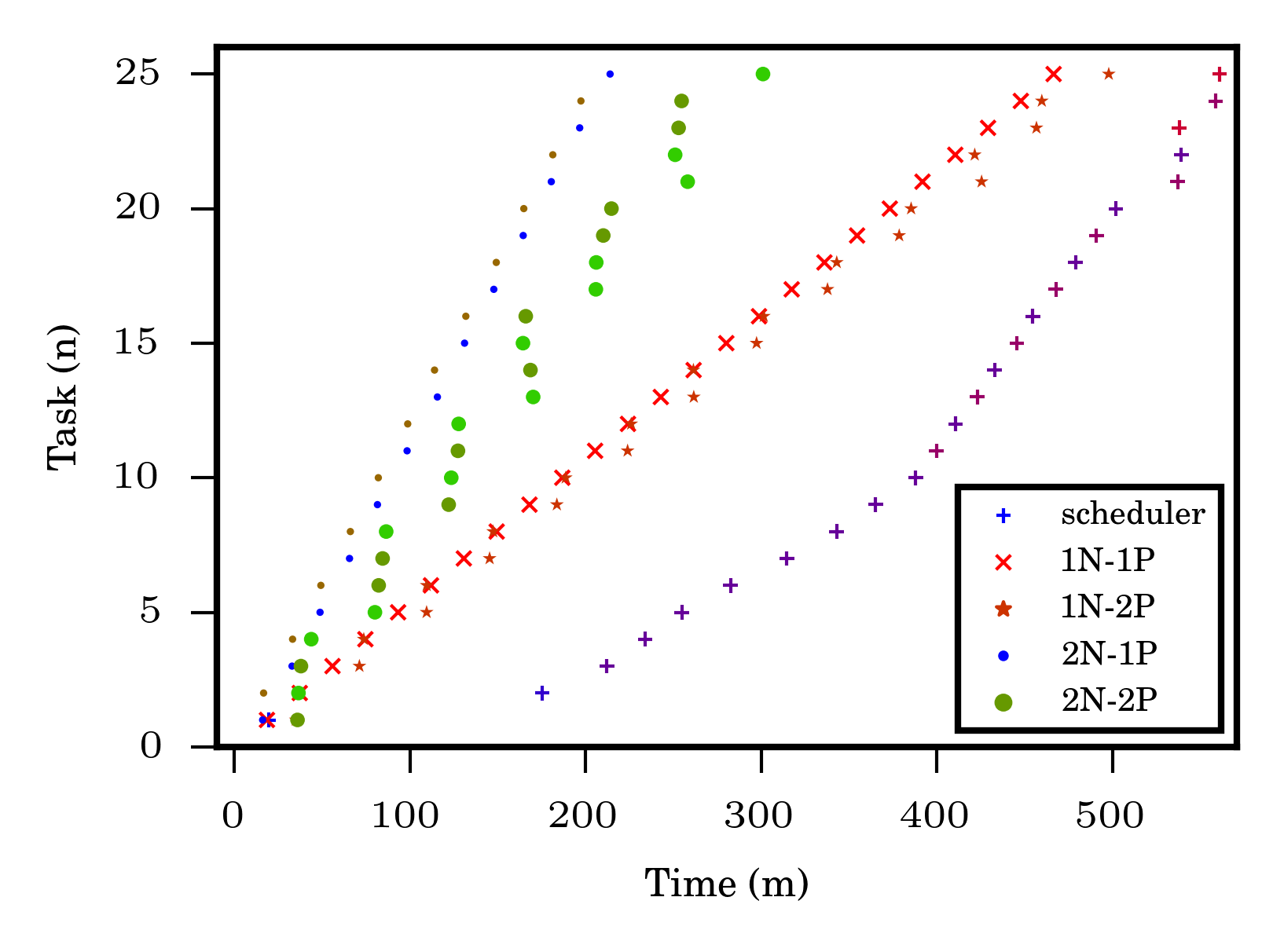}
    \caption{Final execution behavior of NetLogo simulations from Figure~\ref{fig:netlogo_begin}. Each simulation's total execution time was approximately 30 minutes and the cluster's utilization was always above 70\%. \papas\ technique of grouping jobs in MPI-supported clusters is closer to the {\itshape optimal} case (see Figure~\ref{fig:cluster_stats}) while the scheduler operates in the {\itshape normal} regime.}
    \label{fig:netlogo_end}
\end{figure}

In this example, we used the UTK's Advanced Computing Facility (ACF) cluster to implement a parameter
sweep of a NetLogo Behavioral model.
A Lustre parallel file system and the PBS batch system are available.
The model simulates the transmission of Clostridium difficile in a healthcare setting and explicitly incorporates healthcare workers as vectors of transmission while tracking individual patient antibiotic histories and contamination levels of ward rooms due to C. difficile.
We used \papas\ to deploy multiple instances of NetLogo by
varying some XML elements from the original input file.
Input files that were exactly the same for each workflow instance
were placed in a NFS directory, so only a single copy of each was made.
Figures~\ref{fig:netlogo_begin} and~\ref{fig:netlogo_end} show the scheduling and runtime
results of 25 models with varying number of compute nodes per job (N) and number of MPI processes per job (P).
The best results correspond to the scheme that clustered jobs concurrently in multiple
nodes (\textit{i.e.,} 2N-1P and 2N-2P), since
the overall completion time was lowest as well as the number of scheduler interactions.
On the other hand, the worst scheme resulted from submitting jobs independently and
letting the cluster scheduler manage all the jobs.

\section{Performance study: Local matrix multiply application using OpenMP} \label{study_matmul}

Many times software developers need to profile different algorithms, environment
configurations, and parameters for decision-making in terms of algorithms,
hardware, etc.
One such example of a performance study is performing weak and
strong scaling studies at the processor level. OpenMP is a common library
for enabling multiple threads in compute intensive code regions.
Scaling studies run programs by varying the number of OpenMP threads
and the input size. It is common to control the threading configuration via
OpenMP environment variables.
\papas\ is suitable for such scenarios as it allows expressing such experiments
fairly concise.
For example, consider an OpenMP-based matrix multiply application called
{\itshape matmul} that multiplies a pair of randomly generated squared matrices
and has two positional command line options: (1) matrix size and (2) file for resulting matrix.
Let us show how to configure a \papas\ scaling study by running {\itshape matmul}
for input sizes 16--16384 using multiples of 2, and varying the number of OpenMP
threads from 1--8 in steps of 1.
This study corresponds to 88 independent executions of {\itshape matmul}.
Since \papas\ measures the runtime of each task, the application is not mandated
to have an internal timer (unless higher precision is needed).
Additional profiling statistics are up to the user's applications.
Figure~\ref{fig:matmul_params} shows a \papas\ parameter file adhering to the specifications
of the scaling study. Figure~\ref{fig:matmul_graph} shows all the workflow instances
generated for the {\itshape matmul} application.

\begin{figure}
\begin{lstlisting}[frame=single, xleftmargin=0.07\linewidth, linewidth=1.0\linewidth, language={}, basicstyle=\ttfamily\footnotesize, numbers=left, stepnumber=1, keywordstyle=\bf, morekeywords={environ, command}, breaklines=true]
matmulOMP:
    name: Matrix multiply scaling study with OpenMP
    environ:
        OMP_NUM_THREADS:
            - 1:8
    args:
        size:
            - 16:*2:16384
    command: matmul ${args:size} result_${args:size}N_${environ:OMP_NUM_THREADS}T.txt
\end{lstlisting}
    \caption{Example of a \papas\ workflow configuration using YAML for an OpenMP-based
             matrix multiply application. The study performs tasks for both
             weak and strong scaling. Matrix size is varied by doubling
             and number of threads is varied in steps of 1.
             \papas\ keywords are shown in boldface.}
    \label{fig:matmul_params}
\end{figure}

\begin{figure*}
    \includegraphics[scale=0.4]{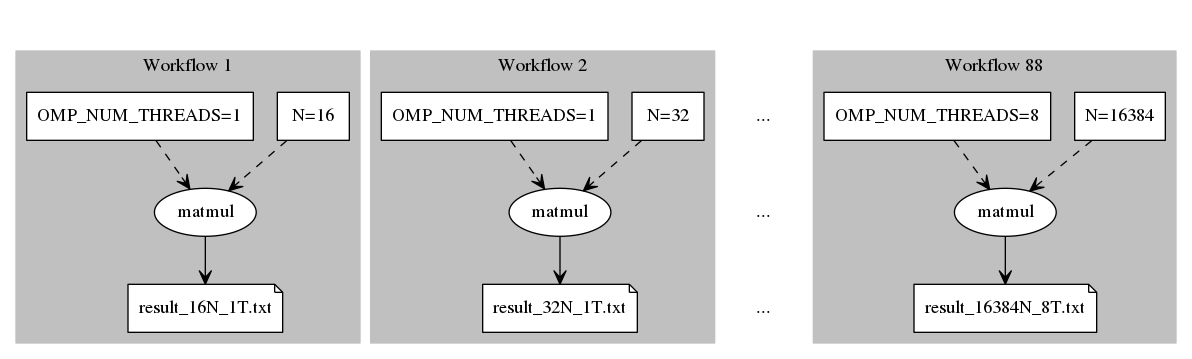}
    \caption{Set of workflow instances generated by \papas\ {\itshape matmul}
             parameter file from Figure~\ref{fig:matmul_params}.}
    \label{fig:matmul_graph}
\end{figure*}

\section{Conclusions} \label{conclusion}

Parameter sweep applications and benchmark applications are common use case examples of scenarios in which scientists seek to identify sets of suitable input parameters and perform application execution.
This work present \papas, a Python 3 generic framework for creating parameter
studies in local, distributed (MPI or SSH), and shared (PBS) computer systems.
A study can consist of a single task, multiple independent tasks, or multiple dependent
tasks (i.e., workflows).
A parameter study can be described using one or multiple simple,
but powerful, parameter files, thus
allowing greater flexibility than most existing tools.
Moreover, \papas\ supports parameters consisting of environment variables,
command line options, files as a whole, file contents, or a combination of these.
\papas\ provides mechanisms to express ranges, Cartesian product, bijection,
and constant parameters, thus, enabling a wide range of possibilities for the user.
Each unique combination of parameters triggers a workflow instance which is executed
independently of other workflow instances.
\papas\ orchestrates the execution of workflow instances, measures runtimes,
and provides the user with provenance information.
By providing these capabilities, \papas\ promises to enhance computational and data science productivity.

\section{Future Work} \label{future}

The \papas\ framework provides exciting support for computational and data science users to achieve higher productivity.
Despite its capabilities, there are numerous extensions to \papas\ under consideration to provide even more usability, flexibility, and productivity.
Future efforts are to integrate \papas\ workflows into grid workflow systems,
such as Taverna and Pegasus, to readily extend the potential \papas\ user community.
One potential approach is to allow the exchange of \papas\ task description files with
Pegasus and similar actively developed workflow management systems.
A \papas\ task internal representation can be converted to define a Pegasus workflow
via the Pegasus Python libraries for writing direct acyclic graphs in XML (DAX).
In this scheme, \papas\ would serve as a
front-end tool for defining parameter studies while leveraging a wide array of
features provided by the Pegasus framework.

Currently, the \papas\ design does not supports nor provides a mechanism to express
automatic aggregation of files, even if tasks utilize the same names for output
files.
Some difficulties that arise with automatic aggregation of files are content ordering
and parsing tasks correctly
(replicated file names). In order to support automatic aggregation,
additional keywords will need to be included in the \papas\ workflow language.

An additional feature to aid in workflow creation is to use a graphical interface
from which the user can define parameter studies. This extension can be designed
with capabilities to create, modify, and/or remove tasks from workflows, as well
as for viewing workflow graphs.

Although there are tools that support inline Python code as the commands
to be executed~\cite{koster2012snakemake}, this ability is constrained from
\papas\ as workflow configuration files are limited by design.

The \papas\ framework will be extended to support tools for measuring application
performance, in addition to the current runtime measures. One popular example of such tools
is PAPI~\cite{terpstra2010papi}.
The current design only measures the runtime of each parameter study workflow,
workflow instance, and task. Higher-detail of profiling metrics could be useful
for: (1) providing the user with additional profiling information, mainly
for benchmarking studies, and (2) as
feedback for improving workflow planning and scheduling decisions.

There is still work to investigate for managing and scheduling parameter
workflows. For example, consider a parameter workflow containing tasks with
same parameters and tasks with multi-valued parameters.
Then, the user may wish to dictate that the set of
workflows will follow a depth-first or breadth-first execution.

These kinds of additional features could significantly broaden the usefulness and resultant productivity improvements provided by \papas.

\begin{acks}

This research is based upon work supported in part by
\grantsponsor{DE-AC05-00OR22725}{UT-Battelle, LLC}{}
under Contract No:{DE-AC05-00OR22725},~\grantnum{DE-AC05-00OR22725},
a joint program with the U.S. Department of Veterans Affairs
under the Million Veteran Project Computational Health
Analytics for Medical Precision to Improve Outcomes Now (MVP-CHAMPION),
and by the joint
\grantsponsor{R01GM113239}{DMS/NIGMS Mathematical Biology
Program} through NIH award No:{R01GM113239}~\grantnum{R01GM113239}.

\end{acks}

\balance
\bibliographystyle{ACM-Reference-Format}
\bibliography{papas}

\end{document}